\documentclass[aps,showpacs,twocolumn,aps,amsmath,amssymb,PRB]{revtex4}
\usepackage{dcolumn}
\usepackage{bm}
\usepackage[dvips]{graphicx}
\usepackage{wrapfig,subfigure}
\usepackage{amssymb}
\usepackage{color}
\usepackage{ulem}

\begin{document}

\title{Mass-loading induced dephasing in nanomechanical resonators}
\author{Juan Atalaya}
\affiliation{Department of Applied Physics, Chalmers University of Technology, G{\"o}teborg Sweden, SE-412 96}

\begin{abstract}
I study dephasing of an underdamped nanomechanical resonator subject to random mass loading of small particles. I present a frequency noise model which describes dephasing due to attachment and detachment of particles at random points and particle diffusion along the resonator. This situation is commonly encountered in current mass measurement experiments using NEM resonators. I discuss the conditions which can lead to inhomogeneous broadening and fine structure in the modes absorption spectra. I also show that the spectra of the higher order cumulants of the (complex) vibrational mode amplitude are sensitive to the parameters characterizing the frequency noise process. Hence, measurement of these cumulants can provide information not only about the mass but also about other parameters of the particles (diffusion coefficient and attachment-detachment rates.)  
\end{abstract}
\date{\today}
\maketitle
\section{Introduction}
It has been demonstrated that nanoelectromechanical (NEM) resonators are very suitable for sensor technology. In particular, it has been shown that 
NEM resonators can be employed to measure mass ~\cite{Ekinci2004, Manalis2007, Zettl2008,Naik2009,Manalis2010} and charge~\cite{Cleland1998, Steele2009,Bachtold2009} with very high precision. In the case of mass measurement applications, NEM resonators are useful because of the combination of properties such as low mass density, high vibrational frequencies and low intrinsic losses. These features are present, {e.g.}, in carbon nanotube based resonators where vibrational frequencies of $f_0\sim 4$~GHz and quality factors of~$Q\sim 10^5$ have been achieved~\cite{Steele2009b,Bachtold2011,Steele2012}. 

I consider an underdamped oscillator whose eigenfrequency exhibits small fluctuations. The eigenfrequency noise, $\Xi(t)$, leads to (classical) dephasing of the oscillator response. Dephasing can be observed, e.g., in a ring down experiment, where, for a given realization of $\Xi(t)$, the oscillations of the oscillator coordinate, $q(t)$, are dephased with respect to the oscillator response in the absence of frequency noise. Moreover, the average oscillator response, $\langle q(t)\rangle_{\Xi(t)}$, decays at a rate which is larger than half the oscillator energy relaxation rate, $2\Gamma$. 

The effect of dephasing can be also observed in the power spectrum or absorption spectrum of the underdamped oscillator as broadening and even departure from the Lorentzian absorption line~\cite{Dykman2010,Atalaya2011a,Dykman2011}. This broadening is not a consequence of an increase of the mechanical losses. Studying frequency noise in nanomechanical resonators is important not only to find cures for the spectral broadening in high quality factor nanoresonators~\cite{Tang2012} but also in the application of the latter as sensors ({e.g.,} NEM-based mass measurement applications)~\cite{Bachtold2012}. 

The eigenfrequency noise may have different origins. For instance, in an electrical read-out method of the nanobeam displacement, charge fluctuations in the nanoresonator cause frequency fluctuations due to capacitive coupling with gate electrodes~\cite{Steele2009,Bachtold2009}. Also, random frequency modulation of a certain vibrational mode can result from parametric coupling with other modes, which are, {e.g.}, thermally driven~\cite{Atalaya2010,Westra2010}. 

In the context of NEM-based mass measurement experiments, particles enter and leave the resonator at random points and they also diffuse along the resonator~\cite{Bachtold2012}. Since the eigenfrequency shift of a vibrational mode depends on the particle position, random mass-loading of particles has been usually perceived as a limiting factor to the precision of NEM-based mass sensors. However, I show that higher order cumulants of the (complex) vibrational mode amplitude can be used to characterize the frequency noise process~\cite{Dykman2011} and determine the particle parameters (mass, diffusion coefficient, attachment and detachment rates.)

In order to quantify the effect of random mass-loading in the average response of nanomechanical resonators, I present a frequency noise model which describes the quasicontinuous type of frequency noise present in NEM-based mass sensors. This model contains,  previously studied, frequency noise models for particle diffusion~\cite{Atalaya2011a} and discrete frequency jumps~\cite{Dykman2010} as limiting cases. 

Bachtold et al. have recently realized CNT-based mass sensors where Xe atoms and Naphthalene molecules attach, detach and diffuse along a CNT nanoresonator~\cite{Bachtold2012}. In typical NEM-based mass measurement experiments the incoming flux rate of particles can be controlled but the initial adsorption point and particle position on the resonator afterwards are beyond experimental control. 

This paper is organized as follows. In section II, I describe the frequency noise model used to calculate the mode susceptibility and higher order cumulants of the complex vibrational amplitude. In sections III and IV, I discuss dephasing due to attachment, detachment and diffusion of a single particle and many particles, respectively. In section V, I summarize the main results of this paper. 

\section{Model}
It has been demonstrated that eigenmodes of nanomechanical resonators can exhibit low intrinsic losses with quality factors $Q\sim10^5$~\cite{Craighead2006,Steele2009b}.  The vibrational amplitude, $q(t)$, of certain eigenmode, which is weakly driven near resonance by a force  $F\cos(\omega_F t)$, satisfies the following equation,
\begin{equation}
\ddot{q} + 2\Gamma \dot{q} + \big[\omega_0^2 + 2\omega_0\Xi(t)\big]q =\frac{F}{M}\cos(\omega_F t) + \frac{f_T(t)}{M},
\label{eq:EOM_1}
\end{equation}
where $\omega_0$, $M$ and $2\Gamma$ are the bare resonance frequency, effective mass and energy relaxation rate of the driven mode, respectively. Equation~\eqref{eq:EOM_1} also includes the eigenfrequency noise, $\Xi(t)$, and the zero-mean additive thermal noise, $f_T(t)$. It is assumed that $\omega_0\gg \Gamma, \Delta\equiv\big(\langle\Xi(t)^2\rangle - \langle\Xi(t)\rangle^2\big)^{1/2}, |\delta\omega|$, $t_c^{-1}$; where $\delta\omega=\omega_F-\omega_0$ is the frequency detuning and $t_c$ is the frequency noise correlation time. It is also assumed that the Fourier components of $\xi(t)$ at frequencies $\omega\sim 2\omega_0$ are negligibly small (no parametric resonance). The relative size between the quantities $\Delta$, $\Gamma$, $t_c^{-1}$ and $|\delta\omega|$ is arbitrary. 

If the above restrictions hold then we can use the rotating-wave-approximation (RWA) to study the dynamics of $q(t)$ at frequencies $\omega\approx\omega_0$. I introduce the slow (complex) dynamical variables $u(t)$ and $u^*(t)$, defined by,
\begin{equation}
{u}(t)=\exp(-i\omega_Ft)\big[i\omega_F q(t)+\dot{q}(t)\big]/2i\omega_F,
\end{equation}
and $q(t) = {u}(t)\exp(i\omega_F t) + {u}^*(t)\exp(-i\omega_F t)$. After averaging out fast oscillating terms, within the oscillation period $2\pi/\omega_F$, the equation of motion for $u(t)$ is~\cite{Dykman1984}  
\begin{equation}
\dot{{u}}(t) = -\big[\Gamma +i(\delta \omega -\Xi(t)) \big]{u} -i\frac{F}{4M\omega_F} +\tilde{f}_T(t), \label{eq:EOM_2}
\end{equation}
where  $\tilde{f}_T(t)=f_T(t)\exp(-i\omega_Ft)/2i\omega_FM$. The value of the moments of the complex vibrational amplitude, $\langle {u}^n(t)\rangle_{st}$, do not depend on the additive thermal noise, $f_T(t)$, if the latter is uncorrelated with the frequency noise process, $\Xi(t)$, and the RWA holds [Eq.~\eqref{eq:EOM_2}].

The frequency noise becomes correlated with the additive thermal noise process in the case that the former depends on the oscillator state (backaction). For instance, backaction occurs in the diffusion-induced bistability mechanism of driven nanomechanical resonators, discussed in Ref.~\cite{Atalaya2011b}, and in nonlinear oscillators, where the vibrational frequency depends on the vibrational amplitude. Below, I neglect these extra mechanisms of dephasing (it is assumed that the vibrational mode is weakly driven.) Thus, I drop $f_T(t)$ from Eq.~\eqref{eq:EOM_2}. 

The frequency noise process, $\Xi(t)=\Xi(\xi(t))$, is assumed to be defined in terms of a Markovian stochastic process $\xi(t)$, which is modeled by the master equation 
\begin{equation}
\partial_t p(\xi,t) = \sum_{\xi'}\hat{W}({\xi,\xi'})p(\xi',t).\label{eq:p_eq}
\end{equation}
Above, $p(\xi,t)$ is the probability distribution of $\xi$ at time $t$ and $\hat{W}(\xi,\xi')$ is the matrix of transition probabilities per unit of time between different $\xi$-states. It is assumed that $\hat{W}$ does not depend on the oscillator state, $u(t)$. Thus, $f_T(t)$ and $\Xi(t)$ are  independent processes. 

The evolution of the joint probability distribution of the oscillator state and the eigenfrequency noise process, $P$, is described by the Einstein-Fokker-Planck equation~\cite{Risken}, 
\begin{eqnarray}
\partial_tP(u,u^*,\xi,t) &=& \hat{W}P + \partial_u\big([\Gamma + i(\delta\omega - \Xi(\xi))]uP\big)  \nonumber \\
&& + \frac{iF}{4M\omega_F} \partial_uP+ \textrm{c.c.}, \label{eq:P_equ}
\end{eqnarray}
where $\Xi(\xi)$ is a scalar function of the noise state $\xi$.

The stationary value of the $n$th moment of the complex vibrational amplitude $u(t)$ is given by,
\begin{eqnarray}
\label{eq:Moments}
 \langle {u}^n(t)\rangle_{st} &=& \sum_{\xi}\iint\textrm{d}u\,\textrm{d}u^*\;{u}^n P_{st}(u,u^*,\xi) ,\nonumber \\
&\equiv& \Big(\frac{F}{4M\omega_F}\Big)^n\chi^{(n)}(\delta\omega),
\end{eqnarray}
where $P_{st}(u,u^*,\xi)$ is the stationary solution of Eq.~\eqref{eq:P_equ} and $\chi^{(n)}(\delta\omega)$ is the scaled $n$th moment of $u(t)$. 

The scaled $n$th moment, $\chi^{(n)}(\delta\omega)$, is written as a sum of scaled partial moments, $\tilde{\chi}^{(n)}(\xi;\delta\omega)$, defined by,
\begin{eqnarray}
\tilde{\chi}^{(n)}(\xi;\delta\omega) &=& \Big(\frac{4M\omega_F}{F}\Big)^{n}\iint\textrm{d}u\,\textrm{d}u^*\;{u}^n P_{st}(u,u^*,\xi), \nonumber \\
\chi^{(n)}(\delta\omega) &=& \sum_{\xi}\tilde{\chi}^{(n)}(\xi;\delta\omega) .
\end{eqnarray}
The scaled partial moments are complex quantities which are coupled by the equations~\cite{Dykman2011}
\begin{equation}
\hat{W}\tilde{\chi}^{(n)} - n\big(\Lambda-i\,\Xi(\xi)\big)\tilde{\chi}^{(n)} = ni\tilde{\chi}^{(n-1)}, \label{eq:chi_n}
\end{equation}
where $\Lambda=\Gamma+i\delta\omega$, $\tilde{\chi}^{(0)}(\delta\omega,\xi)=p_{st}(\xi)$ and $p_{st}(\xi)$ is the stationary solution of Eq.~\eqref{eq:p_eq}. 

Similarly, the (scaled) oscillator susceptibility, $\chi(\delta\omega)$, is defined by
\begin{equation}
\chi(\delta\omega) = \big(4M\omega_F/F\big)\langle u^*(t) \rangle_{st}.
\end{equation}
The scaled oscillator susceptibility can also be written in terms of scaled partial susceptibilities $\chi(\xi;\delta\omega)$ [$\chi(\delta\omega)=\sum_{\xi}\chi(\xi;\delta\omega)$], which are coupled by the equation that results from complex conjugation of Eq.~\eqref{eq:chi_n} with $n=1$. 

The limit cases of slow [$t_c \gg \Gamma^{-1},\Delta^{-1}$] and fast [$t_c \ll \Gamma^{-1},\Delta^{-1}$] frequency noise have been discussed previously~\cite{Dykman2010,Atalaya2011a}. Both limit cases are described by Eq.~\eqref{eq:chi_n} and they correspond to slow (fast) diffusion and  rare (frequent) adsorption and desorption events. In what follows, I discuss the experimentally relevant case of rare adsorption and desorption events, and slow and fast diffusion~\cite{Bachtold2012}. 

The eigenfrequency noise due to random mass-loading in NEM resonators is quasicontinuous. It evolves in time either continuously, when particles diffuse along the nanoresonator, or discontinuously, when particles enter and leave the nanoresonator. Now, I describe a model for the frequency noise when a single particle acts on the resonator. If many particles act independently on the resonator then it is only necessary to model the frequency noise process due to a single particle (cf. appendix B). This frequency noise model includes the frequency noise models for particle diffusion~\cite{Atalaya2011a} and discrete frequency jumps~\cite{Dykman2010} as limit cases.  

The state of the particle is described by two variables $\eta(t)\in\{0,1\}$ and $x(t)\in [-L/2,L/2]$, where $L$ is the resonator length. The former variable has the value of one (zero) when the particle is (is not) on the nanoresonator. The position of the particle on the nanoresonator at time $t$ is $x(t)$. The frequency noise $\Xi(t)$ in Eq.~\eqref{eq:EOM_1} is defined in terms of the variables $x(t)$ and $\eta(t)$ by the relation
\begin{equation}
\Xi\big(\xi\equiv(x,\eta)\big)=-\nu\eta\phi^2(x), \label{eq:xi_one}
\end{equation} 
where $\phi(x)$ is the spatial profile of the driven vibrational mode, $\nu=m\omega_0/2M$, $m$ is the analyte mass, $M= \int\textrm{d}x\, \rho_0 \phi(x)^2$ is the vibrational mode effective mass and $\rho_0$ is the resonator bare mass density. 

The transition probability matrix $\hat{W}(x,\eta;x',\eta')$ for the frequency noise process of a single particle is defined by
\begin{eqnarray}
\partial_tp_{\eta=1}(x,t) &=& -\Gamma_o(x) p_1(x,t)  +  \Gamma_i(x)f(x) p_0(t) \nonumber  \\ 
&& + \hat{L}_D(x)p_1(x,t),  \label{eq:EOM_state_probability}  \\ 
\partial_tp_{\eta=0}(t) &=&  \int_{-L/2}^{L/2} \textrm{d}x\, [\Gamma_o(x) p_1(x,t) - \Gamma_i(x) f(x)p_0(t)], \nonumber 
\end{eqnarray}
where $p_{1}(x,t)$ is the probability density that the particle is on the nanoresonator at position $x$ and time $t$. Similarly, $p_0(t)$ is the probability that the particle is not on the nanoresonator at instant $t$.  The incoming flux distribution per particle is $f(x)$, which is normalized as $\int\textrm{d}x\, f(x)=1$, and $\Gamma_{i(o)}(x)$ is the probability per unit of time that the particle is adsorbed (desorbed) at the point $x$. The detachment rate follows the Arrhenius law, $\Gamma_o\sim \exp(-E_a/k_BT)$, where $k_BT$ is the thermal energy and $E_a$ is the activation energy~\cite{Bachtold2012}. The attachment rate is proportional to the incoming flux of particles and to the cross-section of the nanoresonator. For small vibration amplitudes, the attachment and detachment rates do not depend on the oscillator state, $u(t)$.    

The operator $\hat{L}_D(x)$ describes the diffusion of an overdamped particle,
\begin{equation}
\hat{L}_Dp_1(x,t) = \partial_x(U'(x)p_1)/\kappa m + D\partial_x^2p_1,\label{eq:LD}
\end{equation}
where $\kappa$ and $D$ are the particle friction and diffusion coefficients, respectively, and $U(x)$ is a confining potential acting on the particle. I also assume reflecting boundary conditions:  $U'(x)p_1(x,t)/\kappa m + D\partial_x p_1(x,t)=0$ at ${x=\pm L/2}$. In thermal equilibrium, the diffusion coefficient is determined by the coupling between the particle and the nanobeam vibrational modes (phonon bath) and by temperature. If the excited vibrational mode is of low order and weakly driven, the nanobeam temperature and the other (not directly driven) modes are not significantly affected by the external drive. Hence, for underdamped vibrational modes, it is a good approximation to consider that the diffusion coefficient independent of the vibrational mode state, $u(t)$.

In order that the RWA is valid, the noise correlation time, $t_c\sim\min\{L^2/D,\Gamma_i^{-1},\Gamma_o^{-1}\}$, should be much larger than the oscillation period, $2\pi/\omega_0$. 

Note also that Eq.~\eqref{eq:EOM_state_probability} can be extended to include other subsystems. For instance,  the substrate may be considered as a new subsystem which exchanges particles with the resonator subsystem through the clamping regions. In this case additional exchange rates $\Gamma_{i/o}(x)$ need to be provided. 

\begin{figure}[t]
\includegraphics[width=8.0truecm,trim=1.8cm 3.8cm 3cm 3cm, clip=true]{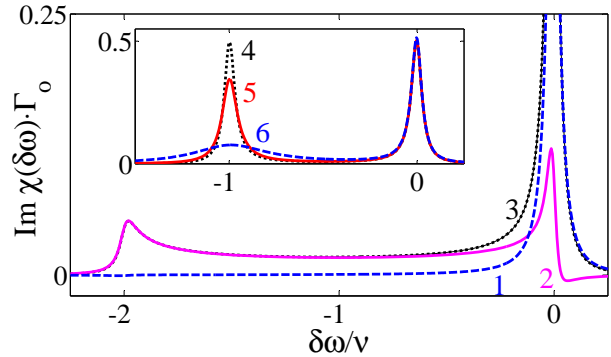}
\centering
\caption{Absorption spectrum, $\textrm{Im} \chi(\delta\omega)$, of the fundamental vibrational mode of a nanobeam subject to   random mass loading from a single particle. Curves 1 (dashed) and 2 (solid) depict the partial susceptibilities $\textrm{Im}\tilde{\chi}_0(\delta\omega)$ and $\textrm{Im}\int\textrm{d}x\,\tilde{\chi}_1(\delta\omega,x)$, and curve 3 (dotted) depicts  $\textrm{Im}\chi(\delta\omega)$ for the case of a nondiffusing particle ($D=0$). {Inset:} Curve 4 (dotted) depicts $\textrm{Im}\chi(\delta\omega)$ for the limit case of an infinitely fast diffusing particle (telegraph noise) and curves 5 and 6 depict $\textrm{Im}\chi(\delta\omega)$ for diffusion coefficients $D=25\Gamma_oL^2$ and $D=2\Gamma_oL^2$, respectively.  Adsorption and desorption rates are $\Gamma_i=\Gamma_o=\nu/30$ and $2\nu$ is the frequency shift due to a particle located at the mode antinode. The oscillator damping rate is $\Gamma=0$. 
\label{fig:fig1}}
\end{figure}
\section{Dephasing from mass loading of a single particle}
 I use the model introduced in section II for the frequency noise to calculate the oscillator susceptibility and also the $n$th moment of the complex vibrational amplitude, $u(t)$. I consider the fundamental flexural mode of a one dimensional resonator with mode shape $\phi(x)=\sqrt{2}\cos(\pi x/L)$. The shift of the vibrational frequency due to a particle at position $x$ is  $-\nu\phi^2(x)$. I assume that the particle may be adsorbed at any point on the resonator with equal probability; {i.e.,} $f(x)=1/L$ in Eq.~\eqref{eq:EOM_state_probability}. Also, the adsorbed particle freely  diffuses along the resonator [$U=0$ in Eq.~\eqref{eq:LD}] with reflecting boundary conditions, $\partial_x p_1(x=\pm L/2,t)=0$. Below, I discuss the case where the maximum frequency shift per particle, $2\nu$, is larger than the oscillator damping rate, $\Gamma$, and the adsorption rate, $\Gamma_i$, and desorption rate $\Gamma_o$~\cite{Bachtold2012}. I discuss first the results for the oscillator susceptibility. 

As discussed in section II, the scaled oscillator susceptibility, $\chi(\delta\omega)$, is given by
\begin{equation}
\chi(\delta\omega) =  \tilde{\chi}_0(\delta\omega) + \int_{-L/2}^{L/2}\textrm{d}x\,\tilde{\chi}_1(\delta\omega,x),
\end{equation}
where $\tilde{\chi}_0(\delta\omega)$ and $\tilde{\chi}_1(\delta\omega,x)$ are the scaled partial susceptibilities associated with no particle and one particle at position $x$ on the resonator. The equation for the scaled partial susceptibilities [cf.~Eq.~\eqref{eq:Eq_partial_suscep}] can be solved analytically in terms of continued fractions (cf.~appendix A).  

\begin{figure}[t]
\centering
\includegraphics[width=\linewidth,trim=0.3cm 3.35cm 2.4cm 2.5cm, clip=true]{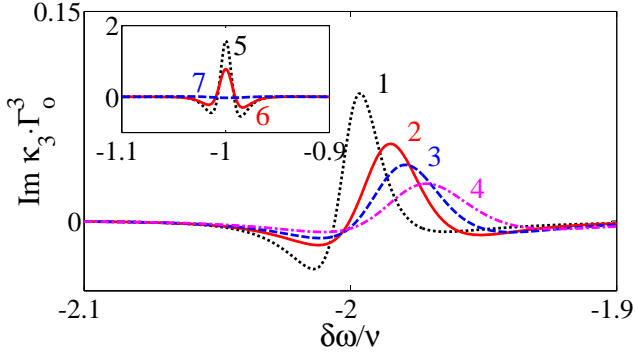}
\caption{Effect of diffusion on the imaginary part of the third cumulant,  $\textrm{Im} \,\kappa_3$, of the complex vibrational amplitude $u(t)$. Curve 1 (dotted) depicts the limit case of a nondiffusing particle ($D=0$). Curves 2 (solid), 3 (dashed) and 4 (dash-dotted) depict the results for a particle with diffusion coefficients $D=10^{-3}\cdot\Gamma_oL^2, 2\cdot10^{-3}\Gamma_oL^2$ and $4\cdot10^{-3}\Gamma_oL^2$, respectively. {Inset}: Curve 5 (dotted) depicts the limit case of an infinitely fast diffusing particle (telegraph noise). Curves 6 (solid) and 7 (dashed) depict the results for $D=250\Gamma_oL^2$  and $D=20\Gamma_oL^2$, respectively. Other parameters: $\Gamma_i=\Gamma_o=\nu/30$ and $\Gamma=0$. 
\label{fig:fig2}}
\end{figure}

Figure~\ref{fig:fig1} depicts the scaled absorption spectrum, $\textrm{Im}\chi(\delta\omega)$, for the limit cases of a slow and fast diffusing particle and $\nu\gg\Gamma_{i},\Gamma_{o},\Gamma$. In the slow diffusion limit, the absorption spectrum exhibits inhomogeneous broadening. For frequency detuning $\delta\omega\approx -2\nu$, the scaled partial susceptibility of a particle at position $x$ is $\tilde{\chi}_1(\delta\omega,x)\approx ip_{1st}(x)/\big[\Gamma +\Gamma_o -i(\delta\omega + \nu\phi^2(x)) \big]$ and the scaled oscillator susceptibility is~\cite{Atalaya2011a}
\begin{equation}
\chi(\delta\omega)\approx \frac{iLp_{1,st}}{\sqrt{\big(\Gamma + \Gamma_o - i\delta\omega \big)\big(\Gamma +\Gamma_o - i(\delta\omega +2\nu)\big)}}. \label{eq:Chi_approx}
\end{equation} 
The scaled partial susceptibility $\tilde{\chi}_0(\delta\omega)$ contributes to the total oscillator susceptibility mainly for frequency detuning $\delta\omega\approx 0$. This is reasonable because $\tilde{\chi}_0(\delta\omega)$ corresponds to no adsorbed particle (zero frequency shift). As depicted in Fig.~\ref{fig:fig1},  $\textrm{Im}\tilde{\chi}_0(\delta\omega)$ is approximately Lorentzian with a half-width $\approx \Gamma+\Gamma_i$ and height proportional the stationary probability of no particle on the resonator: $ p_{0,st}=\Gamma_o/(\Gamma_i+\Gamma_o)$. The asymmetry observed in $\textrm{Im}\tilde{\chi}_0(\delta\omega)$ at  $\delta\omega=0$ is the result of coupling with the scaled partial susceptibilities $\tilde{\chi}_1(x,\delta\omega\approx 0)$.

The limit case of a fast diffusing particle is depicted in the inset of Fig.~\ref{fig:fig1}, assuming that $\nu\gg\Gamma_{i},\Gamma_{o},\Gamma$. In this limit, the oscillator absorption spectrum  can be approximated by two Lorentzians centered at $\delta\omega=0$ and $\delta\omega= -\nu\langle\phi(x)^2\rangle_{p_{1st}}$. These Lorentzians have half-widths approximately equal to $\Gamma+\Gamma_i$ and $\Gamma+\Gamma_o + \nu^2L^2/8\pi^2D$, and heights  approximately  equal to  $p_{0,st}/(\Gamma+\Gamma_i)$ and (L$p_{1,st})/(\Gamma+\Gamma_o+\nu^2L^2/8\pi^2D)$; respectively. These results agree with dephasing due to telegraph frequency noise, which is obtained in the limit  $D\rightarrow\infty$~\cite{Dykman2011}. Thus, the effect of a finite but large diffusion coefficient ($D\gg L^2\Gamma_o$) is to increase the half-width of the Lorentzian centered at $\delta\omega=-\nu$ by an amount $\approx \nu^2L^2/8\pi^2D$. As the particle diffusion gets slower, this Lorentzian gets broader and it eventually looses its Lorentzian shape and acquires the shape given by Eq.~\eqref{eq:Chi_approx}. 

\begin{figure}[t]
\includegraphics[angle=-90,width=.94\linewidth,trim=2.3cm 1.5cm 3.1cm 2.5cm, clip=true]{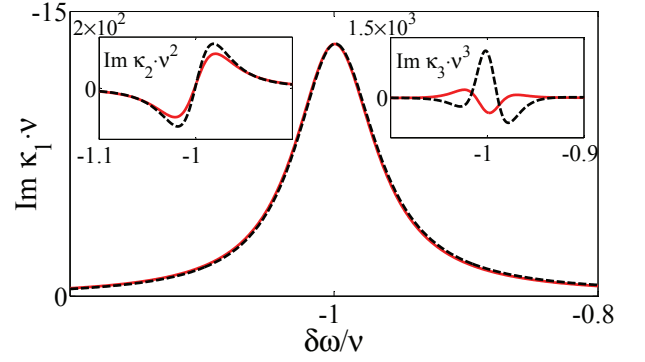}
\centering
\caption{Sensitivity of the cumulants, $\kappa_1$, $\kappa_2$ and $\kappa_3$ of $u(t)$ on the particle diffusion coefficient, $D$.  The dashed black and solid red curves corresponds to $D=3.50\nu L^2$ and $D=3.33\nu L^2$, respectively. The other parameters are derived from $\Gamma+\Gamma_i=3.333\times10^{-2}\nu$, $\Gamma+\Gamma_o + \nu^2L^2/8\pi^2D =3.733\times10^{-2}\nu$, $p_{0,st}=0.505$ and $\nu$ is kept fixed. Note that this choice of parameters slightly affects the first cumulant, $\kappa_1$. Thus, in order to determine all the oscillator and particle parameters, it is necessary to consider the higher order cumulants.
 \label{fig:fig3}}
\end{figure}

Higher order moments of $u(t)$ are calculated by solving numerically the coupled system~\eqref{eq:chi_n}. In the characterization of the dephasing process, cumulants are more important than the moments of $u(t)$ because the former vanish in the absence of  frequency noise~\cite{Dykman2011}. Figure~\ref{fig:fig2} shows the spectrum of the imaginary part of the third cumulant, $\kappa_3 = \langle u^3\rangle - 3\langle u\rangle\cdot\langle u^2\rangle + 2\langle u\rangle^3$, for both limit cases of a fast and a slow diffusing particle. The magnitude and shape of the higher order cumulants spectra depend on the dephasing process parameters (i.e., diffusion coefficient and attachment and detachment rates); moreover, this dependency is more significant for higher order cumulants of $u(t)$. 

Using a frequency noise model such as the one introduced in section II,  measurement of the cumulants of $u(t)$ can provide enough information to characterize the dephasing process and determine the particle parameters. This is demonstrated in Fig.~\ref{fig:fig3} where  the imaginary parts of the second and third cumulant are shown to be more sensitive to a small change of the diffusion coefficient (solid and dashed lines correspond to $D=3.33L^2\nu$ and $D=3.50L^2\nu$, respectively.) Note that the other particle parameters and oscillator damping rate have been chosen such that the first cumulant is minimally affected (i.e., $\Gamma+\Gamma_i$, $\Gamma+\Gamma_o+\nu^2L^2/8\pi^2D$, $\nu$ and $p_{0,st}$ are fixed in the solid and dashed curves of Fig.~\ref{fig:fig3}). 

In order to measure the moments and cumulants of the slowly varying complex amplitude $u(t)\equiv X(t)-iY(t)$, one has to measure the inphase, $X(t)$, and quadrature, $Y(t)$, envelopes of the oscillations: $q(t)=2X(t)\cos(\omega_Ft) +2Y(t)\sin(\omega_Ft)$. In carbon-based nanoresonators, these measurements can be performed using the FM mixing technique~\cite{Ayari2010}.  I point out that the recording time of $X(t)$ and $Y(t)$ has to be much larger than the correlation time of the frequency noise and the oscillator energy relaxation time. 

\section{Dephasing from mass loading of many particles}
In this section I consider dephasing due to particles acting independently on the nanoresonator. Particles act independently on the resonator if the average particle density on the resonator is small such that they do not interact with each other. For simplicity, it is assumed that the particles have identical mass,  adsorption and desorption rates and diffusion coefficient. Hence, the total frequency noise $\xi(t)$ is equal to a sum of independent and identical processes $\xi_i(t)$, which are realizations of the process~\eqref{eq:xi_one},  
\begin{equation}
\xi(t)=\sum_{i=1}^{N} \xi_{i}(t), \label{eq:xi_many_particles}
\end{equation}
where $N$ is the total number of particles in the system. Particles can be either in the gas subsystem or in the resonator subsystem. The number of particles on the resonator $n(t)$ evolves according to a Bernoulli process with master equation
\begin{eqnarray}
\dot{p}_{n} &=& -[\Gamma_on +\Gamma_i(N-n)\big]p_n + (n+1)\Gamma_op_{n+1} + \nonumber\\
&&\Gamma_i\big(N-(n-1)\big)p_{n-1}, \label{eq:Bernoulli}
\end{eqnarray}
where $N_0=N\Gamma_i/(\Gamma_i+\Gamma_o)$ is the average number of particles on the resonator. The stationary distribution of Eq.~\eqref{eq:Bernoulli} for $N_0\ll N$  and $\Gamma_i=(N_0/N)\Gamma_o\ll\Gamma_o$ is the Poisson distribution: $p^{st}_n=\exp(-N_0)N_0^n/n!$. 

\begin{figure}
\includegraphics[angle=-90,width=0.9\linewidth,trim=1cm 1.5cm 3.3cm 3cm, clip=true]{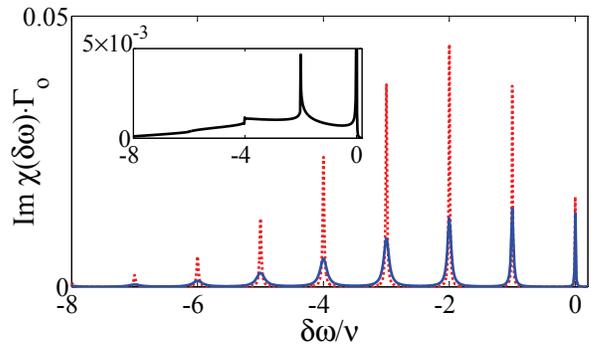}
\centering
\caption{Absorption spectrum, $\textrm{Im}\chi(\delta\omega)$, of the fundamental flexural mode of a nanobeam subject to random mass loading of many particles. The average number of particles on the resonator is $N_0=3$. The dotted curve depicts the results for the limit case of infinitely fast diffusing particles ($D\rightarrow\infty$). The solid line depicts the result for diffusion coefficient $D=500\Gamma_oL^2$. {Inset:} Limit case of slowly diffusing particles ($D\rightarrow 0$). Other parameters: $N=10^4$, $\Gamma_o=0.002\nu$,  $\Gamma_i=(N_o/N)\Gamma_o$ and $\Gamma=0$. 
 \label{fig:fig4}}
\end{figure}

Figure~\ref{fig:fig4} depicts the oscillator absorption spectrum for  $N_0=3$ and $\nu\gg\Gamma_o,\Gamma$. The limit of fast diffusing particles agrees with the results obtained previously for discrete frequency jumps with a stationary Poisson distribution. The absorption spectrum displays a fine structure (dotted curve) and it is formed by Lorentzians centered at frequencies $\delta\omega=-j\nu$ ($j=0,1,\cdots$) with half-widths $\Gamma_j=\Gamma+(j+N_0)\Gamma_o$ [$(j+N_0)\Gamma_o $ is equal to the reciprocal mean time of having $j$ particles on the resonator], and the heights are proportional to the stationary distribution $p^{st}_j$~\cite{Dykman2010}. 

The effect of a finite diffusion coefficient, $D$, is that the half-width of the above Lorentzians is increased by an amount $\approx j\nu^2L^2/(8\pi^2D)$, cf. Fig.~\ref{fig:fig4} (solid curve). In the limit of slow diffusion, I again observe inhomogeneous  broadening due to the randomness associated with the particle insertion point. This is depicted in the inset of Fig.~\ref{fig:fig4}. Here, the absorption spectrum, $\textrm{Im}\chi(\delta\omega)$, also exhibits sharp peaks. However, these peaks are not Lorentzians and they get sharper as $\Gamma_o$ decreases. They also appear in the case of dephasing due to a single particle acting on the resonator, cf. Eq.~\eqref{eq:Chi_approx}. The above results agree with Monte Carlo simulations~\cite{Gillespie77,Mannella2000}.  

\section{Conclusions}
I have discussed the effect of random mass loading of small particles on the  resonant response of the fundamental flexural mode of a nanobeam. A model for the frequency noise has been presented which accounts for particle attachment and detachment at random points on the resonator and particle diffusion along the resonator. Analytical and numerical results have been presented for the mode susceptibility and for some higher order cumulants of the complex vibrational mode amplitude. In the case of rare adsorption and desorption events and fast diffusion, the oscillator absorption spectrum exhibits fine structure. The diffusion contribution to the linewidths of the (approximately) Lorentzian lines, centered at $-j\nu$, is $\approx j\nu^2L^2/(8\pi^2D)$, where $D$ is the diffusion coefficient, $j=(0,1,\cdots)$, $2\nu$ is the frequency shift due to a particle located at the vibrational mode antinode and $L$ is the nanobeam length. It has been also demonstrated that higher order cumulants are more sensitive to the parameters of the frequency noise and their measurement can be used to develop a mass sensing scheme, which is able not only to determine the mass of the particle but also its diffusion coefficient, attachment and detachment rates to the nanoresonator.


\section{Appendix A}
Here I derive the analytical expression for the (scaled) oscillator susceptibility $\chi(\delta\omega)=(4M\omega_F/F)\langle u^*(t)\rangle$ for the case of mass loading due to a single particle, cf. section III. The assumptions are uniform incoming flux distribution $f(x)=1/L$ and free diffusion along the resonator [$U(x)=0$]. The equation for the scaled susceptibility is given by the complex conjugate of Eq.~\eqref{eq:chi_n} with $n=1$ and $\hat{W}$ defined by the master equation~\eqref{eq:EOM_state_probability}, whose stationary solution is $p_{0,st}=\Gamma_o/(\Gamma_i+\Gamma_o)$ and $p_{1,st}(x)=(1/L)\cdot\Gamma_i/(\Gamma_i+\Gamma_o)$. The total scaled susceptibility is equal to the sum of the scaled partial susceptibilities $\tilde{\chi}_0(\delta\omega)$ (no particle on resonator) and $\tilde{\chi}_1(x;\delta\omega)$ (particle at position $x$),
\begin{equation}
\chi(\delta\omega) =  \tilde{\chi}_0(\delta\omega) + \int_{-L/2}^{L/2}\textrm{d}x\,\tilde{\chi}_1(x;\delta\omega),
\end{equation}
and the scaled partial susceptibilities satisfy the equations
\begin{eqnarray}
\big[\Lambda^*-i\nu\phi^2(x)+\Gamma_o- D\partial_x^2\big]\tilde{\chi}_1 -\Gamma_i f(x) \tilde{\chi}_0 &=& ip_{1,st.}(x), \nonumber \\
\big(\Lambda^* + \Gamma_i\big)\tilde{\chi}_0 - \Gamma_0 \int \textrm{d}x\, \tilde{\chi}_1(\delta\omega,x)  &=& ip_{0,st.}, \label{eq:Eq_partial_suscep}
\end{eqnarray}
where $\Lambda^*=\Gamma - i\delta\omega$. The solution for $\tilde{\chi}_1(x;\delta\omega)$ is sought in the form  
\[
\tilde{\chi}_1(x;\delta\omega) = \sum_{k\geq0} A_k\cos(2\pi k x/L).
\]

The coefficients $A_k$ satisfy a difference equation system which can be solved in terms of continued fractions,
\begin{eqnarray}
\int\textrm{d}x\, \tilde{\chi}_1(x;\delta\omega) &=& \frac{ L\cdot i p_{1,st.} + i\Gamma_{i}p_{0,st.}/\big(\Gamma + \Gamma_i  - i\delta\omega\big)}{R(D,\delta\omega) - \Gamma_i\Gamma_o/\big(\Gamma+\Gamma_i-i\delta\omega\big) },\nonumber\\
\tilde{\chi}_0(\delta\omega)  &=& \frac{\Gamma_o\int\textrm{d}x\, \tilde{\chi}_{1}(\delta\omega,x) + ip_{0,\,st.}}{\Gamma +\Gamma_i - i\delta\omega}, \label{eq:sol_1particle_part2}
\end{eqnarray}
where $R$ is given by 
\begin{equation}
R(D,\delta\omega) = \Theta(0) + \cfrac{\nu^2/2}{\Theta(1)+\cfrac{\nu^2/4}{\Theta(2) + \cfrac{\nu^2/4}{\ddots}}}, \label{eq:cfrac}
\end{equation}
and $\Theta(n)=\Gamma + \Gamma_o - i(\delta\omega +\nu) + n^2\tau_D^{-1}$. The correlation time of frequency noise due to only particle diffusion is $\tau_D=(L/2\pi)^2/D$~\cite{Atalaya2011a}.  

\section{Appendix B}
Here I consider the calculation of the (scaled) susceptibility function, $\chi(\delta\omega)$, for the case of many particles acting independently on the resonator, cf. section IV. The scaled oscillator susceptibility can be expressed as 
\begin{equation}
\chi(\delta\omega) = i\int_{-\infty}^{0} \textrm{d}t\, e^{\Lambda^*t}\cdot g^N(t),
\label{eq:chi_many_particles}
\end{equation}
where $N$ is the total number of particles in the system (in and out of the resonator), $\Lambda^*=\Gamma-i\delta\omega$ and the dephasing-dependent term $g(t)$ is equal to 
\begin{equation}
g(t) = \int\textrm{d}x\, \check{\tilde{\chi}}_1(x,t) + \check{\tilde{\chi}}_0(t).
\end{equation} 
The time-dependent partial susceptibilities $\check{\tilde{\chi}}_1(x,t)$ and $\check{\tilde{\chi}}_0(t)$ are related to $\tilde{\chi}_1(x;\delta\omega)$ and $\tilde{\chi}_0(\delta\omega)$ by Fourier transform 
\begin{equation}
\tilde{\chi}_1(x;\delta\omega) = \int_{-\infty}^0 \textrm{d}t\, e^{\Lambda^*t} \check{\tilde{\chi}}_1(x,t),
\end{equation}
and a similar relation between $\check{\tilde{\chi}}_0(t)$ and $\tilde{\chi}_0(\delta\omega)$ holds. These new (scaled) partial susceptibilities satisfy the equations,
\begin{eqnarray}
\partial_t\check{\tilde{\chi}}_1 &=& \big[\Gamma_0 -i\nu\phi(x)^2 -D\partial_x^2\big]\check{\tilde{\chi}}_1 -\Gamma_if(x)\check{\tilde{\chi}}_0,\nonumber\\
\partial_t\check{\tilde{\chi}}_0 &=& \Gamma_i\check{\tilde{\chi}}_0 -\Gamma_o\int\textrm{d}x\,\check{\tilde{\chi}}_1(x,t),
\end{eqnarray}
with initial conditions $\check{\tilde{\chi}}_1(x,0)=ip_{1st}(x)=i(1/L)\cdot\Gamma_i/(\Gamma_i+\Gamma_o)$ and $\check{\tilde{\chi}}_0(0)=ip_{0st}=i\Gamma_o/(\Gamma_i+\Gamma_o)$ with reflecting boundary conditions: $\partial_x\check{\tilde{\chi}}_1(x=\pm L/2,t)=0$. Note that $g(t)$ can be also obtained by taking inverse Fourier transform of the result \eqref{eq:sol_1particle_part2} with $\Gamma=0$. 

\section{Acknowledgments}
This work was supported in part by the Swedish VR and SSF
and by the EC project QNEMS (FP7-ICT-233952). I also acknowledge Dr. Andreas Isacsson for helpful discussions.  

\bibliographystyle{apsrev}

\end{document}